# Micropolarity-Ramification of Laminar/Turbulent Circular-Plane-Jet

Abuzar A. Siddiqui[†]

*Department of Basic Sciences, Bahauddin Zakariya University, Multan, 60800, Pakistan.*
*(abuzarabid@bzu.edu.pk)*

In the present work we formulated the boundary-value-problem, comprising partial differential equations (PDEs) of steady flow for laminar/turbulent circular jet of a micropolar fluid. A new boundary layer-similarity transformation/solution was derived which is valid not only for the Newtonian fluids but also for the micropolar fluids. Through this transformation PDEs are transformed into the ordinary differential equations (ODEs). These ODEs were solved numerically by the finite-difference method. The obtained results were compared with existing results [9] for the Newtonian fluids. The comparison was favourable. The micropolarity influences were highlighted in the present work. The axial-fluid-speed and normal stress-component $\sigma_{zz}$ decreases but radial-fluid-speed, microspin and normal stress-component $\sigma_{\theta\theta}$ increase as the micropolarity effect enhances.

The fluid-speed, microspin, shear stresses, normal stresses and couple stresses are dominant in the vicinity of the jet-source whereas they all vanish as $r \to \infty$ far away from the jet-source analogous to the plane-jet-flow [8]. In addition, axial-speed, discharge, microspin, shear stresses and the components of couple stress ($m_{r\vartheta}$ and $m_{\vartheta r}$) intensify with $z$ linearly but radial velocity, normal stresses and the component of couple stress $m_{\vartheta z}$ are independent of $z$. The micropolarity-effects on discharge diminish if the diameter of jet-source reduces. Analogous to the plane-jet-flow [8], $tr(\sigma_{lm}) = 0$, for the micropolar as well as for the Newtonian fluids. Both the normal stresses $\sigma_{\theta\theta}$ and $\sigma_{zz}$ decreases as $r$ goes on increasing, analogous to the shear-stresses. But all the normal components of the couple stress tensor vanish. The couple stress tensor is symmetric and skew-symmetric if $\beta \mp \gamma = 0$ analogous to the plane-jet flow [8]. Finally the fluid-speed will also intensify if turbulent-influences are enhanced.

## 1. INTRODUCTION

The Navier-Stokes equations act as the equations of motion for flow of the Newtonian fluids---like alcohol, gasoline, milk, oil, water etc. But the flow of many techno-scientifically indispensable fluids--- like blood, colloidal suspensions, epoxies, liquid crystals, slurries, etc---cannot be elucidated by the Newtonian fluids. This category of fluids comprises colloids/semi-rigid particles that not only translate but also rotate about axes passing through their centroids. These fluids can sustain body forces and usual (Cauchy) stress tensor as simple Newtonian fluids do in addition to the body couple and the couple stress tensor. These fluids exhibit six degree of freedom---three more than the Newtonian fluids. These fluids are named as the Cosserat (micropolar) fluids. Their equations of motion and constitutive equations were derived by Eringen [1] in 1966. These fluids have a lot of applications in the industries such that they are being used on analyzing blood flow [2-3], polymeric suspensions [4], lab-on-a-chip [5-6], rigid-rod epoxies [7] and much more [8].

On motivating the industrial and techno-scientific dominance of the micropolar fluids, we embarked the program to investigate the micropolarity effects on the circular jet. A circular jet is a long narrow circular orifice (analogous to the point source) that can emit the fluid continuously freely and mixes with surrounding-stationary fluid [9]. It gained great attention not only in the previous century like [9, 10] but also it is burning issue of this era [11-12]. It is due to fact that, it has tremendous and marvelous applications in industry and engineering. For example: in cooling hot material like steel plates at ROT [13], in cooling of combustion engines and electronic microchips [14], in setting up the ink-jet printers [15-16], in evaporation of refrigerant-oil mixture [17], in Erosion threshold of a liquid immersed granular bed [18] and much more [10].

The plan of this paper is: Section 2 contains mathematical formulation of the boundary-value-problem for circular jet flow and similarity transformations; the computational procedure will be presented in section 3 while section 4 comprises brief review of the Schlichting's solution for the Newtonian Fluids. The results (in graphical form) in addition to the discussions hereon, are given in section 5 whereas conclusions are made briefly in section 6.

## 2. BASIC ANALYSIS

Let us write the governing equations, neglecting thermal effects, of an incompressible viscous micropolar (Cosserat) fluid [1], in dimensional form, as:

$$\mathbf{V}'_{n,n} = 0, \tag{1}$$

$$\rho \dot{\mathbf{V}}'_k = \rho \mathbf{b}'_k + \boldsymbol{\sigma}'_{jk,j}, \tag{2}$$

and 
$$\rho j_o \dot{\mathbf{N}}'_k = \rho \mathbf{c}'_k + \mathbf{m}'_{jk,j} + \varepsilon_{kjn} \boldsymbol{\sigma}'_{jn}. \tag{3}$$

In addition, its constitutive equations are [8, 19]

$$\boldsymbol{\sigma}'_{kj} = -p\delta_{kj} + (\mu + \chi)(\mathbf{V}'_{j,k} + \mathbf{V}'_{k,j}) + \chi(\mathbf{V}'_{k,j} - \varepsilon_{kjn}\mathbf{N}'_n), \tag{4}$$

and
$$\mathbf{m}'_{kj} = a_1 \mathbf{N}'_{n,n}\delta_{kj} + a_2 \mathbf{N}'_{k,j} + a_3 \mathbf{N}'_{j,k}. \tag{5}$$

Here the superscript dot signifies for the material derivative such that $\mathbf{V}'$ and $\mathbf{N}'$ are, respectively, are the (dimensional) fluid-velocity and the microspin of the colloids in the fluid; $\mathbf{b}'$ and $\mathbf{c}'$ are, respectively, are the body-force and the body-couple; $p, \rho$ and $j_o$ are, respectively, the fluid-pressure, the fluid-density and the local micro (colloidal)-inertia [2]. Apart three spin-gradient-viscosity-coefficients $\{a_1, a_2, a_3\}$ while $\{\lambda, \mu, \chi\}$ are three spin-viscosity-coefficients. These are related as the following traditional form of the Clausius-Duhem inequality, viz.: [19]

$$\left.\begin{array}{l} 3a_1 + a_2 + a_3 \geq 0, \; a_3 \geq a_2, \; a_3 \geq 0 \\ 3\lambda + 2\mu + \chi \geq 0, \; 2\mu \geq -\chi, \; \chi \geq 0 \end{array}\right\} \tag{6}$$

On the basis of Equations (1) – (5), we are interested to investigate the flow-characteristics and micropolarity effects in the flow in a circular jet (laminar as well as turbulent). The circular jet is a long narrow circular orifice with diameter $D$. It can emit the fluid continuously freely and steadily in a stationary fluid [9]. Obviously, the suitable coordinates system will be cylindrical polar coordinates system $(r', \theta, z')$ in which the origin coincides with the centroid of the orifice-aperture such that $z'$ – axis resides along the axial of the jet as shown in the figure 1.

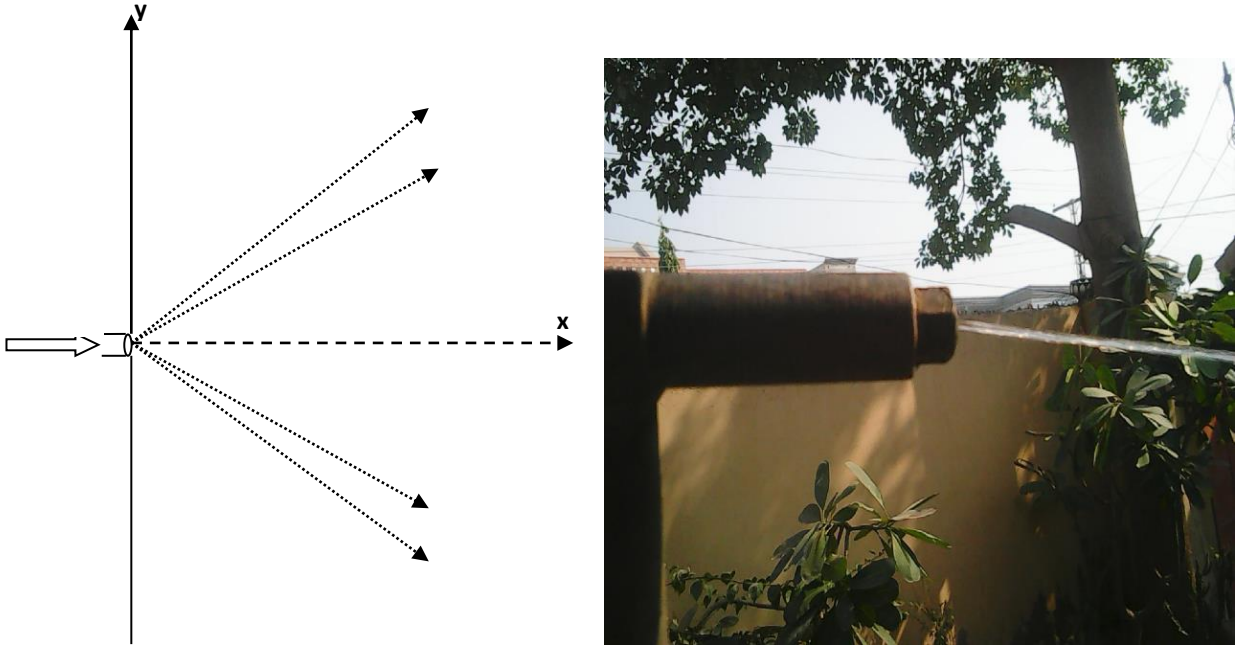

**FIG. 1**. Flow configuration: A circular jet.

The circular-jet-flow is really three dimensional but conveniently we consider it two dimensional in the light of following five assumptions:

- flow is steady and axially-symmetric such that the fluid-velocity and micro-spin vector can be expressed, respectively, as $\mathbf{V}'_i = u'\hat{\mathbf{e}}_r + w'\hat{\mathbf{e}}_z$ and $\mathbf{N}'_i = S\hat{\mathbf{e}}_\theta$;

- all the physical quantities are independent of transverse angle ($\theta$);

- the ambient fluid is not influenced by any external forces like the body-force and the body-couple;

- the constant fluid-pressure is assumed so that the constant radial moment ($M'_r = \rho \int r'w'^2 dr'$) exists; [9]

- all the assumption of Newtonian-Prandtl boundary layer are preserved and applicable for the micropolar (Cosserat) fluid.

So the governing equations (1) – (3), in the light of aforementioned assumptions, in the boundary layer equations form will be:

$$\frac{u'}{r'} + \frac{\partial u'}{\partial r'} + \frac{\partial w'}{\partial z'} = 0, \tag{7}$$

$$\nu_m \left[ \frac{\partial^2 w'}{\partial r'^2} + \frac{1}{r'^2}\frac{\partial w'}{\partial r'} \right] + \frac{\chi}{\rho}\left[ \frac{\partial S'}{\partial r'} + \frac{S'}{r'} \right] = u'\frac{\partial w'}{\partial r'} + w'\frac{\partial w'}{\partial z'}, \tag{8}$$

$$\gamma\left[ \frac{\partial^2 S'}{\partial r'^2} + \frac{1}{r'^2}\frac{\partial S'}{\partial r'} - \frac{S'}{r'^2} \right] - \chi\left[ \frac{\partial w'}{\partial r'} + 2S' \right] = \rho j_o\left[ u'\frac{\partial S'}{\partial r'} + w'\frac{\partial S'}{\partial z'} \right], \tag{9}$$

where $\nu_m = (\mu + \chi)/\rho$.

The boundary conditions will be

$$u' = 0, \qquad w' = w'_m \qquad \text{at} \quad r' = 0, \tag{10}$$

$$\{u', w', S'\} \to \{0,0,0\}. \tag{11}$$

Here $w'_m = \max(w')$. The above boundary conditions are caused by no-slip, and matching conditions.

Next, let us apply the dimension analysis on above boundary-value-problem (BVP), it will deform as:

$$\frac{u}{r} + \frac{\partial u}{\partial r} + \frac{\partial w}{\partial z} = 0, \tag{12}$$

$$\frac{\partial^2 w}{\partial r^2} + \frac{1}{r^2}\frac{\partial w}{\partial r} + K_1\left[\frac{\partial S}{\partial r} + \frac{S}{r}\right] = u\frac{\partial w}{\partial r} + w\frac{\partial w}{\partial z}, \quad (13)$$

$$K_2\left[\frac{\partial^2 S}{\partial r^2} + \frac{1}{r^2}\frac{\partial S}{\partial r} - \frac{S}{r^2}\right] - K_3\left[\frac{\partial w}{\partial r} + 2S\right] = u\frac{\partial S}{\partial r} + w\frac{\partial S}{\partial z}, \quad (14)$$

with subject to the boundary conditions:

$$u = 0, \qquad w = w_m \qquad \text{at} \quad r = 0, \quad (15)$$

$$\{u, w, S\} \to \{0,0,0\}. \quad (16)$$

where $(u, w) = D v_m^{-1}(u', w')$, $(r, z) = D^{-1}(r', z')$ and $S = D^2 v_m^{-1} S'$ such that the coupling parameter, $K_1 = \chi/(\mu+\chi)$, $K_2 = \gamma/(\mu+\chi)j_o$ and $K_3 = (D^2 K_1)/j_o$. Moreover, the stress and couple-stress tensors are made dimensionless as: $\boldsymbol{\sigma}_{kl} = \rho D^2(\mu+\chi)^{-2}\boldsymbol{\sigma}'_{kl}$ and $\mathbf{m}_{kl} = D^3(\gamma v_m)^{-1}\mathbf{m}'_{kl}$.

The aforementioned BVP comprises the coupled non-linear partial differential equations. These can be transformed to set of ordinary differential equations by introducing the following similarity analysis:

$$u(r,z) = -r^{-1}F(r), \quad w(r,z) = zr^{-1}\frac{dF(r)}{dr}, \quad \text{and} \quad S(r,z) = -zr^{-1}G(r); \quad (17)$$

such that continuity equation (11) are automatically satisfied by the solution $(u,w)$ of (17). Furthermore, if (17) is introduced in Eq. (13) and Eq. (14), we get

$$\frac{d^3F(r)}{dr^3} - \frac{1}{r}\frac{d^2F(r)}{dr^2} + \frac{1}{r^2}\frac{dF(r)}{dr} + K_1\left[\frac{G(r)}{r^2} - \frac{1}{r}\frac{dG(r)}{dr}\right] = \frac{F(r)}{r^2}\frac{dF(r)}{dr} - \frac{F(r)}{r}\frac{d^2F(r)}{dr^2} + \frac{1}{r}\left(\frac{dF(r)}{dr}\right)^2, \quad (18)$$

$$K_2\left[\frac{1}{r^2}\frac{d^2G(r)}{dr^2} - \frac{3}{r^3}\frac{dG(r)}{dr} + \frac{3G(r)}{r^4}\right] + \left[\frac{1}{r^3}\frac{dG(r)}{dr} - \frac{2G(r)}{r^4}\right]F(r)$$
$$= K_3\left[\frac{2G(r)}{r^2} + \frac{1}{r^2}\frac{dF(r)}{dr} - \frac{1}{r}\frac{d^2F(r)}{dr^2}\right] + \frac{G(r)}{r^3}\frac{dF(r)}{dr}. \quad (19)$$

The boundary conditions in similarity variable will become as:

$$F(0) = 0, \quad G(0) = \beta_0\left[1 - r\frac{d^2F(0)}{dr^2}\right], \quad \frac{dF(0)}{dr} = 1, \text{ and } F(\infty) = G(\infty) = \frac{dF(\infty)}{dr} = 0.$$

The boundary value problem (BVP) given above contains coupled, non-linear ordinary differential equations and cannot be solved analytically. Therefore, it will be solved by the numerical scheme whose description is given in next section.

## 3. COMPUTATIONAL PROCEDURE

If we take

$$\phi(r) = \frac{dF(r)}{dr} \tag{20}$$

then equations (18) and (19) will take the form as:

$$\frac{d^2\phi(r)}{dr^2} - \frac{1}{r}\frac{d\phi(r)}{dr} + \frac{\phi(r)}{r^2} + K_1\left[\frac{G(r)}{r^2} - \frac{1}{r}\frac{dG(r)}{dr}\right] = \frac{F(r)\phi(r)}{r^2} - \frac{F(r)}{r}\frac{d\phi(r)}{dr} + \frac{(\phi(r))^2}{r}, \tag{21}$$

$$K_2\left[\frac{1}{r^2}\frac{d^2G(r)}{dr^2} - \frac{3}{r^3}\frac{dG(r)}{dr} + \frac{3G(r)}{r^4}\right] + \left[\frac{1}{r^3}\frac{dG(r)}{dr} - \frac{2G(r)}{r^4}\right]F(r) = K_3\left[\frac{2G(r)}{r^2} + \frac{\phi(r)}{r^2} - \frac{1}{r}\frac{d\phi(r)}{dr}\right] + \frac{G(r)\phi(r)}{r^3}. \tag{22}$$

The central finite-difference approximations are applied to the derivatives involved in Eqs. (21) – (22) to get a set of two finite difference equations. These are further solved by the Successive-Over-Relaxation method [21] whereas Eq. (20) will be integrated by the Simpson's rule [21].

## 4. THE SCHLICHTING-SOLUTION AND THE NEWTONIAN FLUIDS

Before discussing the results, let us review and reproduce the existing solution for a Newtonian fluid, which was developed by Schlichting [9] and may be named it as the Schlichting-solution.

If we put $\chi = 0$ (or $S' = 0$) in Eq. (8), we get the boundary layer equation for the Newtonian fluid

$$\nu\left[\frac{\partial^2 w'}{\partial r'^2} + \frac{1}{r'^2}\frac{\partial w'}{\partial r'}\right] = u'\frac{\partial w'}{\partial r'} + w'\frac{\partial w'}{\partial z'} \tag{23}$$

Schlichting considered the following similarity solution [9]:

$$w' = \nu r^{-1}\frac{df}{d\eta} \tag{24}$$

where $\eta = r/z$.

On introducing Eq. (23) in Eq. (24), it can easily be found that:

$$\frac{d^3 f(\eta)}{d\eta^3} - \frac{1}{\eta}\frac{d^2 f(\eta)}{d\eta^2} + \frac{1}{\eta^2}\frac{df(r)}{d\eta} = \frac{f(\eta)}{\eta^2}\frac{df(\eta)}{d\eta} - \frac{f(\eta)}{\eta}\frac{d^2 f(\eta)}{d\eta^2} + \frac{1}{\eta}\left(\frac{df(\eta)}{d\eta}\right)^2$$

This equation was solved analytically with subject to the aforementioned boundary conditions by Schlichting as: [9]

$$w' = \frac{0.375k}{\pi v z}\left[1 + 0.25\xi^2\right]^{-2}, \tag{25}$$

where $\xi = \alpha_1 \eta$ such that $\rho k = M'_r$.

## 5. RESULTS AND DISCUSSIONS

In this section, the graphical touch of the numerical results and discussions thereon are presented in order to highlight the influence of micropolarity on circular (laminar/turbulent) jet. For this, we plotted graphs for physical parameters of fluid such as fluid-speed (radial/axial), fluid-flux, fluid-stresses and colloidal-couple-stresses. The notable thing is that all the numerical calculations are made for fixed material-constants $\mu = 3 \times 10^{-2}$ Pa.s, $\rho = 1.2 \times 10^3$ Kg.m$^{-3}$ [8]. The couple stress parameter $\beta \in [-\gamma, \gamma]$ and viscosity coupling parameter $K_1 \in [0,1)$ are used [8]. Out of [0,1), three values of $K_1$ namely $K_1 = \{0,\ 0.2,\ 0.6\}$ are selected for the display of the axial-fluid-speed in figure 2.

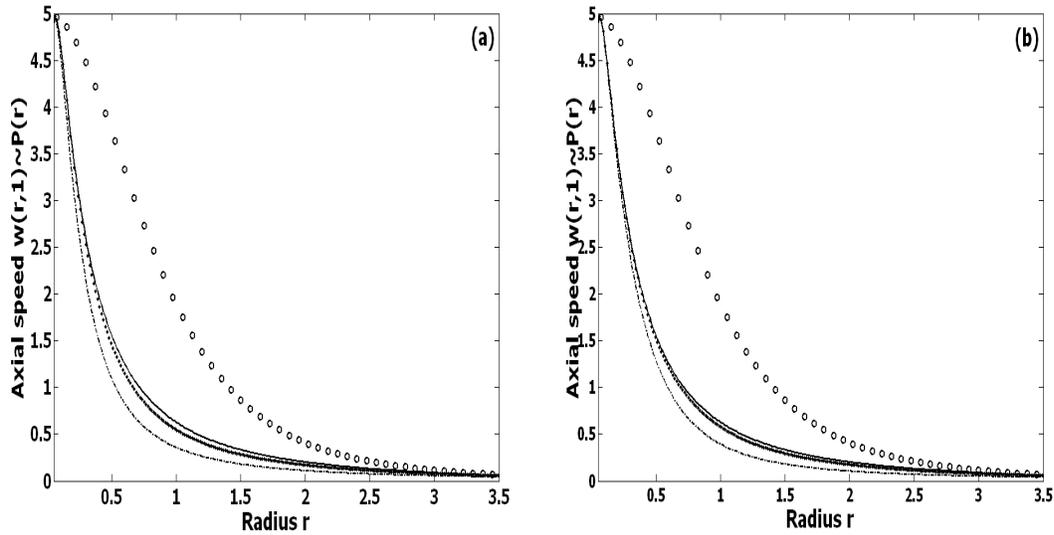

**FIG. 2**. Variation of laminar-axial-speed $w$ with radius $r$ for present result $K_1$=0 (solid curve); $K_1$=0.2 (dotted curve); $K_1$=0.6 (dotted-dashed curve); for (a) $D$=0.1 (b) $D$=0.05. The present results are compared with the results obtained analytical-solution Eq. (24), which is represented here by circle for the Newtonian fluids ($K_1$=0), when $\alpha_1 = 1.58$.

This figure shows that the axial-speed decays with the increase of $K_1$ for all value of $r$, $D$ and $z$. This implies that the axial speed will decrease as the micropolarity effect enhances. In addition, it is not surprising that this axial speed will also decrease with the increase of diameter $D$ of jet for the

Newtonian and micropolar fluids. This decreasing effect is also implementable if we move far away from the jet source consistently for both types of fluids. Like the plane-jet [8], the axial speed $w \to 0$ as $r \to \infty$. It is by virtue of the fact that fluid is stationary far away from the jet source, which is also in justification with boundary condition (16), for the Newtonian as well as for the micropolar fluids. The effect of $z$ on $w$ was analyzed but it is not displayed here. It is observed that $w$ intensifies with z linearly, which is in coincidence with the similarity solution (17). Moreover, figure 2 also depicts the comparison of present results with the existing results [9] for the Newtonian fluids ($K_1 = 0$). Although both the results don't coincide exactly yet their variational-trend confirms our results.

Unlike the axial-fluid-speed, the radial-fluid-speed $u$ increases with the increase of diameter D of jet for all values of *z, r* and $K_1$ (i.e., for the Newtonian as well as micropolar fluids). This effect was highlighted in figure 3. This figure shows that $u$ also increases if the micropolarity-effects increase for all values of *z, r* and *D*. Furthermore, the radial-fluid-speed is independent of z (owing to (17)). It is dominant in the vicinity of the center of the jet and vanishes far away from the jet, analogous to the axial-fluid-speed, for both the fluids.

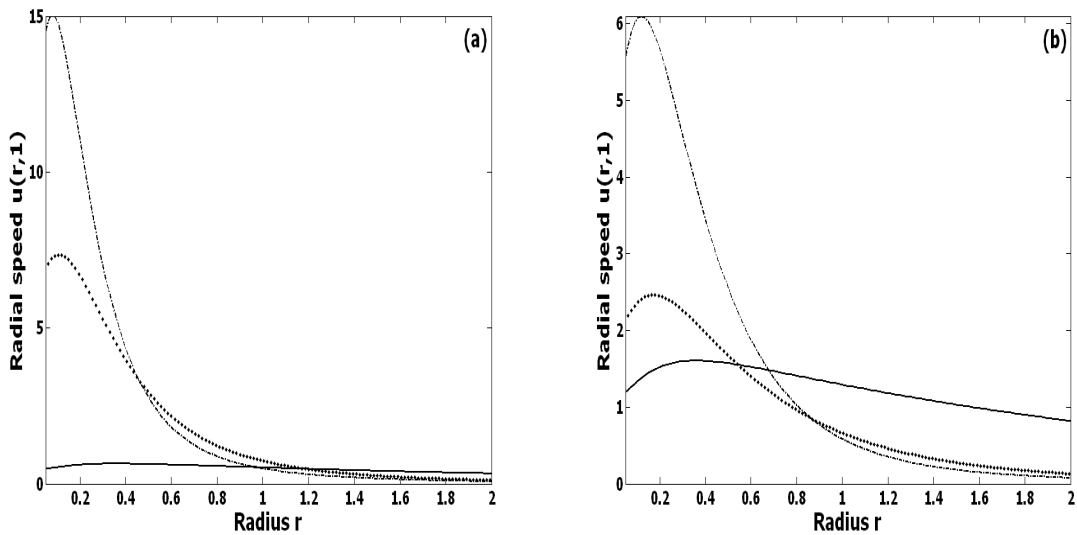

**FIG. 3**. Variation of radial-speed *u* with radius *r* at *z*=1 for $K_1$=0 (solid curve); $K_1$=0.2 (dotted curve) and $K_1$=0.6 (dotted-dashed curve), when (a) *D*=0.1, (b) *D*=0.05.

It will be better to review the fact that the there is no existence of microspin S of aciculate colloides in the Newtonian fluids, it occurs merely in the micropolar fluids. Therefore, we presented the results for S only for micropolar fluids ($K_1 \neq 0$) in figure 4. This figure represents that microspin is also dominant in the vicinity of the jet-source like the fluid speeds. It intensifies with the rise of micropolar effects (i.e., as

$K_1$ increases) for all $z$, $r$ and $D$. This rise in microspin is also observed with $z$, as it is linearly proportional to $z$, for all values of $K_1$, $D$ and $r$. This fact is in accordance with similarity solution (17).

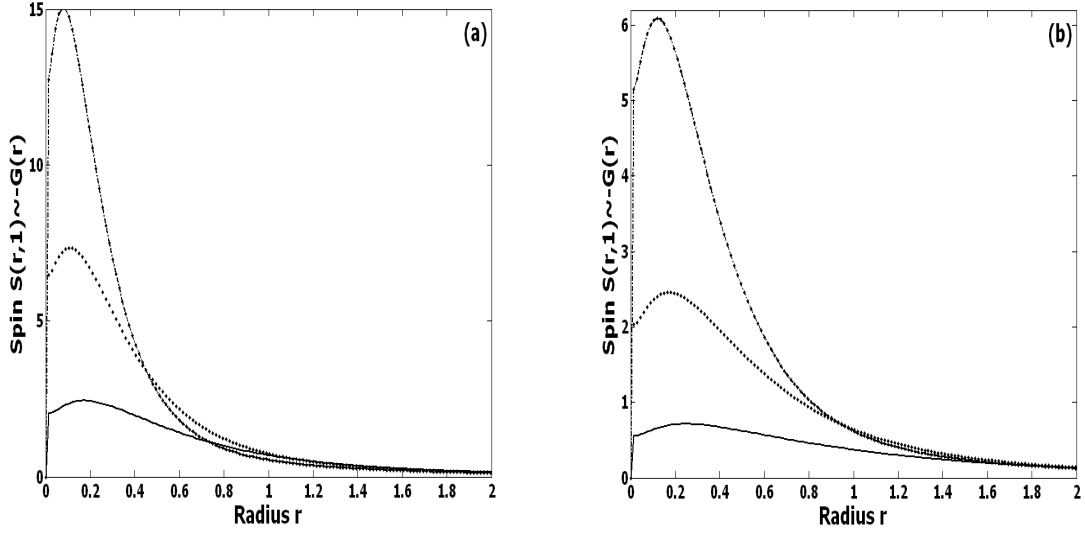

**FIG. 4**. Variation of Microspin $S$ with radius $r$ at $z=1$ for $K_1=0.05$ (solid curve); $K_1=0.2$ (dotted curve) and $K_1=0.6$ (dotted-dashed curve), when (a) $D=0.1$, (b) $D=0.05$.

Now let us probe the micropolarity-effects on discharge (or fluid-flux). It has a vital role in analyzing fluid dynamics, regarding the engineering perspective. It is usually defined as [9]: $Q' = \int_0^\infty r'w'dr'$ such that $Q'$ is related with $Q$ as; $Q' = Dv_m Q$. In the present context, it will take the following simple form as:

$$Q = 2\pi z F(\infty), \qquad (26)$$

by virtue of (17). This Eq. (26) shows that $Q \propto z$ for a fixed value of $K_1$ since $F(\infty)$ varies with $K_1$ irrespective the jet-diameter $D$. Furthermore, it is important to note that the dependency of discharge on $K_1$ diminishes as $D$ reduces. In addition $Q$ rises as $z$ increases (in downstream too) for both the fluids (Newtonian as well as micropolar) as shown in figure (5), which is in agreement with the existing result (for Newtonian fluid) [9]. This effect is very analogous to the flux in plane-jet flow [8].

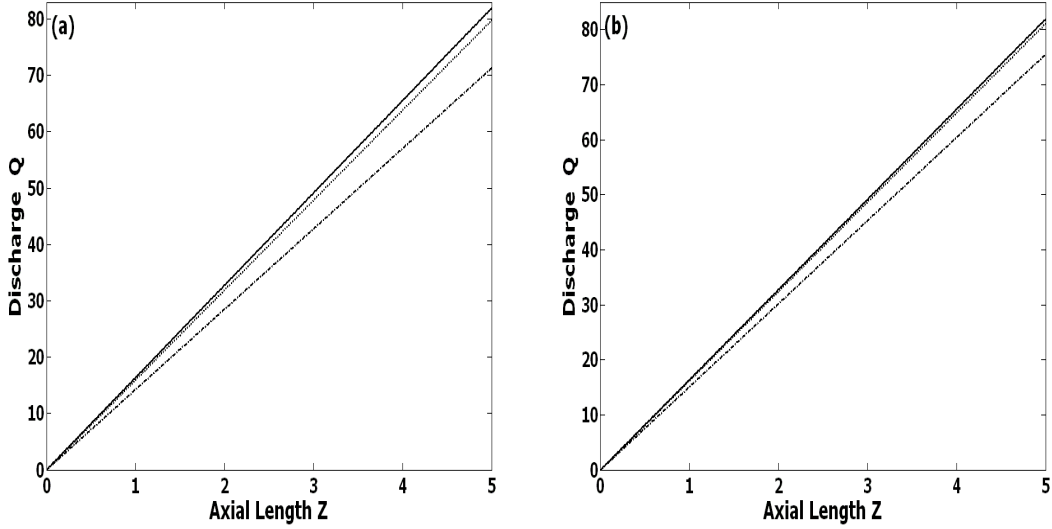

FIG. 5. Variation of discharge/fluid-flux $Q$ with $z$ for $K_1=0$ (solid curve); $K_1=0.01$ (dotted curve) and $K_1=0.05$ (dotted-dashed curve) when (a) $D=0.1$ (b) $D=0.05$.

In the light of aforementioned assumptions of section 2, the non-zero components of dimensionless-stresses are $(\sigma_{rz}, \sigma_{zr}, \sigma_{rr}, \sigma_{\theta\theta}, \sigma_{zz})$-----two shear and three normal stresses. These are related with fluid-speeds and microspin, by virtue of Eq. (4), as:

$$\sigma_{rz} = (1-K_1)\frac{\partial w}{\partial r} + \frac{\partial u}{\partial z} + K_1 S, \tag{27}$$

$$\sigma_{zr} = (1-K_1)\frac{\partial u}{\partial z} + \frac{\partial w}{\partial r} - K_1 S, \tag{28}$$

$$\sigma_{rr} = (2-K_1)\frac{\partial u}{\partial r}, \tag{29}$$

$$\sigma_{\theta\theta} = (2-K_1)\frac{u}{r}, \tag{30}$$

and $$\sigma_{zz} = (2-K_1)\frac{\partial w}{\partial z}. \tag{31}$$

Their variation is examined and is displayed in figures (6) and (7) for selected values of $K_1 = \{0, 0.2, 0.6\}$. These figures depict that, for the Newtonian and micropolar fluids, the following facts are common: (i) both the shear stresses ($|\sigma_{rz}|$ and $|\sigma_{zr}|$) are dominant at and in the vicinity of axis of jet; and (ii) $|\sigma_{rz}|$ decreases while $|\sigma_{zr}|$ increases as $K_1$ increases $\forall r$; but (iii) both $|\sigma_{rz}|$ and $|\sigma_{zr}|$ decrease with the increase of $r$ such they vanishes as $r \to \infty$ (far away from the jet-source). This last fact is in the accordance with boundary condition (16). Furthermore, the shear stresses $\sigma_{lm} \propto z$, $\forall r$ and $\forall K_1$ provided that $l \neq m$. Unlike shear stresses, the normal stresses are independent of $z$ as shown in figure 7. Like the plane-jet-flow [8], $tr(\sigma_{lm}) = 0$, that implies zero influence of pressure, which is in accordance with the aforementioned assumption made in section 2. In addition, we observe that, like shear stresses, the normal stresses $\sigma_{\theta\theta}$ and $\sigma_{zz}$ are significant at and in the vicinity of the $r = 0$ such that $\sigma_{rr} = -(\sigma_{\theta\theta} + \sigma_{zz})$. Furthermore, normal stress $\sigma_{\theta\theta}$ increases if the micropolarity effect increases but it is no so for $\sigma_{zz}$. However, both the normal stresses decreases as $r$ goes on increasing, analogous to the shear-stresses.

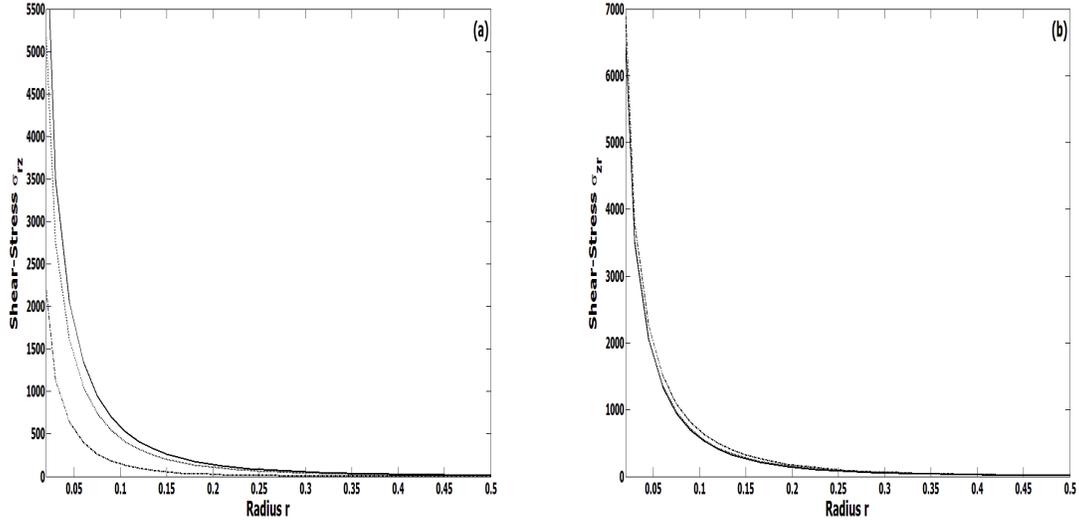

FIG. 6. Variation of Shear-Stresses (a) $\sigma_{rz}$ (b) $\sigma_{zr}$, with $r$ at $z=1$ for $K_1=0$ (solid curve); $K_1=0.2$ (dotted curve) and $K_1=0.6$ (dotted-dashed curve) when $D=0.1$.

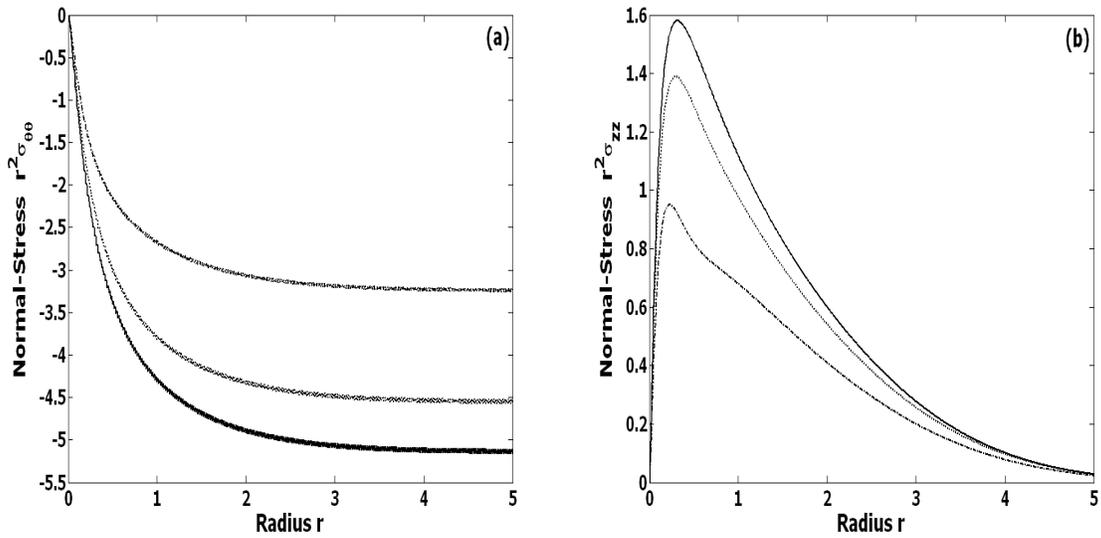

FIG. 7. Variation of normal-stresses (a) $r^2 \sigma_{\theta\theta}$ (b) $r^2 \sigma_{zz}$, with $r$ at $z=1$ for $K_1=0$ (solid curve); $K_1=0.2$ (dotted curve) and $K_1=0.6$ (dotted-dashed curve) when $D=0.1$.

The stress exists in the Newtonian as well as in the micropolar fluids but the couple stress occurs in the micropolar fluids merely. It is caused by the local spin of colloids in the fluids. Generally, the constitutive equation, given in Eq. (5), determines its weight. In our case, this couple stress tensor will have four non-zero components $(m_{r\theta}, m_{\theta r}, m_{rz}, m_{zr})$, which are given explicitly as follows:

$$m_{r\theta} = K_4 \frac{\partial S}{\partial r} - \frac{S}{r}, \tag{32}$$

$$m_{\theta r} = \frac{\partial S}{\partial r} - K_4 \frac{S}{r}, \tag{33}$$

and
$$m_{z\theta} = K_4 m_{\theta z} = \frac{\partial S}{\partial z}. \tag{34}$$

where $K_4 = \frac{\beta}{\gamma} \in [-1,1]$ [25]. These components of the couple stress are plotted in figures (8) and (9) for selected values of parameters as mentioned in the captions. These figures shows that $m_{r\vartheta}$ and $m_{\vartheta r}$ are linearly proportional to $z$ but $m_{\vartheta z}$ is independent of $z$ which is in accordance with the (17). In addition, all the non-zero components of the couple stress are dominant in the vicinity of $r = 0$ and they diminish as $r \to \infty$ far away from the jet-source. Furthermore they intensify as the micropolar influences are increased. In general, there are two notable things. One is that, the all the normal components of the couple stress tensor vanishes for this flow. Other one is that, this couple stress tensor is symmetric and skew-symmetric if $\beta \mp \gamma = 0$ analogous to the plane-jet flow [8].

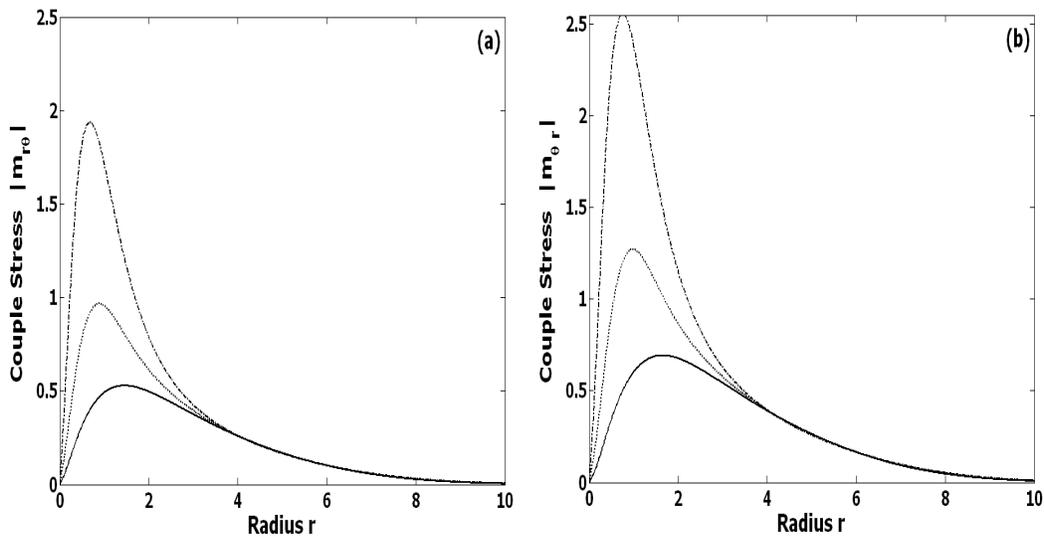

FIG. 8. Variation of couple stress (a) $r^3 |m_{r\theta}|$ (b) $r^3 |m_{\theta r}|$ with $r$ at $z=1$, when $K_4 = 0.5$ (or $2\beta = \gamma$) for $K_1$=0.007 (solid curve); $K_1$=0.02 (dotted curve) and $K_1$=0.06 (dotted-dashed curve).

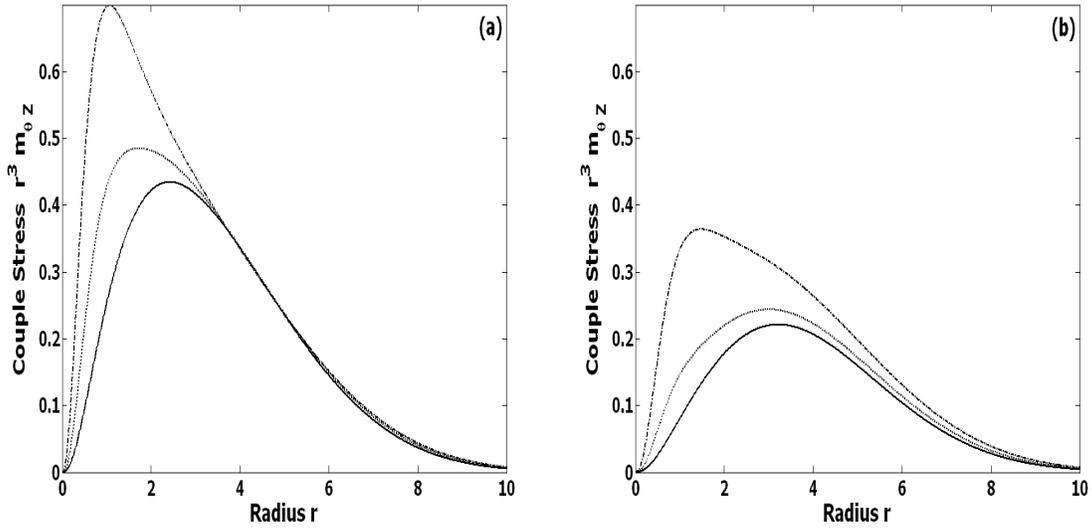

FIG. 9. Variation of couple stress $r^3 |m_{\theta z}|$ with $r$, when $K_4 = 0.5$ (or $2\beta = \gamma$) for $K_1$=0.007 (solid curve); $K_1$=0.02 (dotted curve) and $K_1$=0.06 (dotted-dashed curve) if (a) $D$=0.1 (b) $D$=0.05.

Finally, we examine the turbulent behavior of the circular jet briefly. For this we plotted the dimensional axial-fluid-speed for laminar as well as turbulent flows and are presented here in figure 10. This figure shows the increasing/decreasing trend is common in both laminar and turbulent boundary layers. However the rise in the central-line speed for turbulent flow can easily observed in figure 10(b). Consequently, the fluid-speed will also intensify if turbulent-influences are enhanced.

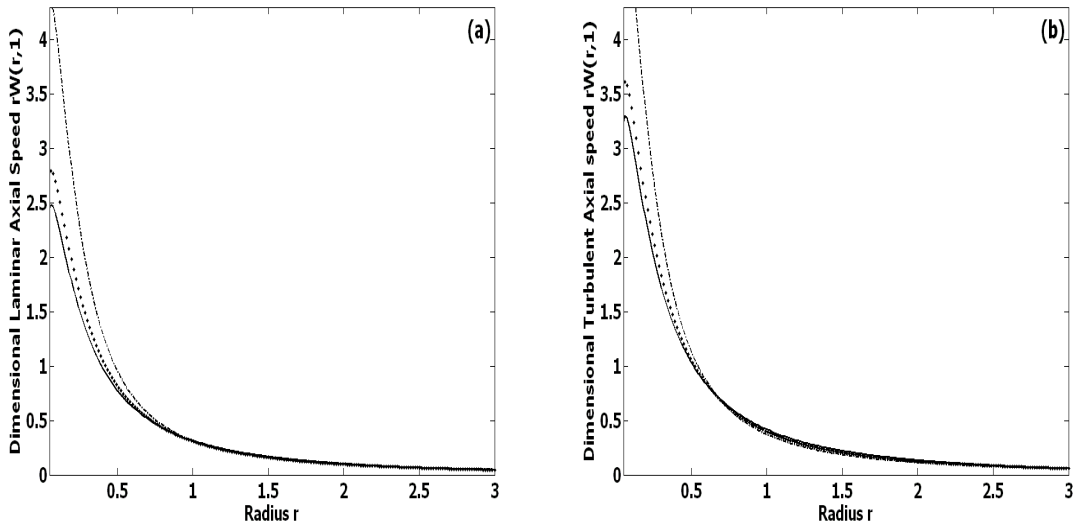

FIG. 10. Variation of dimensional-axial-speed W=w'(r, 1) with radius $r$ when $D$=0.1 for $K_1$=0 (solid curve); $K_1$=0.2 (dotted curve); $K_1$=0.6 (dotted-dashed curve); if flow is (a) laminar (b) turbulent.

## 6. CONCLUSIONS

In the present work we formulated the boundary-value-problem (BVP), comprising partial differential equations (PDEs) of steady flow for laminar/turbulent circular jet of a micropolar fluid. A new boundary layer-similarity transformation/solution was derived which is valid not only for the Newtonian fluids but also for the micropolar fluids. Through this transformation PDEs are transformed into the ordinary differential equations (ODEs). These ODEs were solved numerically by the finite-difference method. The obtained results were compared with existing results [9] for the Newtonian fluids. The comparison was favourable. As the aciculate particles in a micropolar fluid can spin without translation, the micropolarity must have influence on fluid-physical-parameters (e.g. fluid-speed, internal stresses, couple stresses and discharge). This influence was highlighted in the present work. The axial-fluid-speed and normal stress-component $\sigma_{zz}$ decreases but radial-fluid-speed, microspin and normal stress-component $\sigma_{\theta\theta}$ increase as the micropolarity effect enhances (i.e., as $K_1$ increases). Same is happened if diameter $D$ is increased not only for the micropolar but also for the Newtonian fluids. The dependency of discharge on $K_1$ diminishes as $D$ reduces.

The fluid-speed, microspin, shear stresses, normal stresses and couple stresses are dominant in the vicinity of the jet-source whereas they all vanish as $r \to \infty$ far away from the jet-source analogous to the plane-jet-flow [8]. In addition, axial-speed, discharge, microspin, shear stresses and the components of couple stress ($m_{r\vartheta}$ and $m_{\vartheta r}$) intensify with $z$ linearly but radial velocity, normal stresses and the component of couple stress $m_{\vartheta z}$ are independent of $z$. The micropolarity-effects decrease (i.e., dependency of discharge on $K_1$) diminishes if the diameter of jet-source reduces. The shear stress-component $|\sigma_{rz}|$ decreases while another shear stress-component $|\sigma_{zr}|$ increases as micropolar effects goes on increasing ($K_1$ increases) $\forall r$. But both $|\sigma_{rz}|$ and $|\sigma_{zr}|$ decrease with the increase of $r$. Analogous to the plane-jet-flow [8], $tr(\sigma_{lm}) = 0$ for the micropolar as well as for the Newtonian fluids. Both the normal stresses $\sigma_{\theta\theta}$ and $\sigma_{zz}$ decreases as $r$ goes on increasing, analogous to the shear-stresses. All the normal components of the couple stress tensor vanish for this flow. Other one is that, this couple stress tensor is symmetric and skew-symmetric if $\beta \mp \gamma = 0$ analogous to the plane-jet flow [8]. Finally the fluid-speed will also intensify if turbulent-influences are enhanced.